\documentclass[letterpaper,notoc]{CECS}

\preprint{{\tiny CECS-PHY-02/08}\\{\tiny GACG-PHY-12/03} }
\title{Hamiltonian Treatment of the Gravitational Collapse of Thin Shells}
\author{Juan Cris\'{o}stomo\thanks{jcrisost@gacg.cl}\\Instituto deF\'{\i}sica, Facultad de Ciencias B\'{a}sicas y Matem\'{a}ticas,
Pontificia Universidad Cat\'{o}lica de Valpara\'{\i}so, Avenida
Brasil 2950, Valpara\'{\i}so, Chile.\\Centro de Estudios
Cient\'{\i}ficos (CECS), Casilla 1469, Valdivia, Chile.}

\author{Rodrigo Olea\thanks{rolea@fis.puc.cl}\\Pontificia
Universidad Cat\'{o}lica de Chile, Casilla 306, Santiago 22,
Chile.\\Centro de Estudios Cient\'{\i}ficos (CECS), Casilla 1469,
Valdivia, Chile.}

\input epsf

\abstract{A Hamiltonian treatment of the gravitational collapse of
thin shells is presented. The direct integration of the canonical
constraints reproduces the standard shell dynamics for a number of
known cases. The formalism is applied in detail to three
dimensional spacetime and the properties of the (2+1)-dimensional
charged black hole collapse are further elucidated. The procedure
is also extended to deal with rotating solutions in three
dimensions. The general form of the equations providing the shell
dynamics implies the stability of black holes, as they cannot be
converted into naked singularities by any shell collapse
process.}

\begin{document}

\section{Introduction}

The gravitational collapse of thin shells was beautifully discussed in the
classic paper of Israel \cite{cruz-israel}. The generalization to include
electric charge was given by Kucha\v{r} \cite{kuchar}, and an interesting
further development and applications were given by Ipser and Sikivie \cite
{ipser-sikivie}. In all these treatments the analysis is based on the
discontinuities in the intrinsic and extrinsic curvatures of the world tube
of the collapsing matter as he regards it as embedded either in its exterior
or its interior.

In this paper we introduce, in addition to the intrinsic and extrinsic
geometry of the world tube, another structure, namely, a foliation of
spacetime by constant time surfaces which intersect the tube. The reason for
doing this is that the charges of the black hole which results from the
collapse of the shell are conserved quantities given by surface integrals at
spacelike infinity, which are naturally treated in terms of the Hamiltonian
formalism \cite{Regge:1974zd}. Furthermore, the local properties of the
horizon, such as its area, are also economically treated in Hamiltonian
terms. The formalism which emerges from combining the Israel treatment with
the Hamiltonian formalism is quite compact and permits to economically
analyze a number of situations of interest.

The plan of the paper is as follows. Section 2 and Section 3 briefly reviews
the Israel method for thin shell collapse and Hamiltonian formalism,
respectively. Section 4 applies the canonical formalism to the gravitational
collapse of a spherically symmetric shell in an spacetime of arbitrary
dimension and recovers and further clarifies results previously found in the
literature. Section 5 studies the radial gravitational collapse in three
dimensions spacetimes, including the electrically charged case. Section 6
extends the treatment to deal with angular momentum in three-dimensional
spacetimes. Finally, Section 7 is devoted to brief concluding remarks.

\section{Thin Shell Formalism Revisited}

The standard treatment for dealing with thin shells in General Relativity,
arising from the seminal work of Israel \cite{cruz-israel} has provided a
useful tool to tackle a large variety of cases, ranging from lower
dimensional static black hole formation \cite{steif-peleg} to interesting
recent applications in the analysis of the junction conditions for extended
objects in Gauss-Bonnet extended gravity (see, e.g., \cite{EGB}).

The standard procedure considers a timelike hypersurface $\Sigma _{\xi }$,
generated by the time evolution of the shell. This hypersurface divides the
spacetime into two regions, namely, $V_{+}$ and $V_{-}$. Let $\xi ^{\mu }$
the outer pointing, unit normal to the \emph{world tube}, which is
spacelike, and $h_{ab}$ the induced metric on the tube. Here, the indices $%
a,b=1,...,(d-1)$ label the tangent directions along the hypersurface. The
coordinates set $\{x^{\mu }\}$ describes the spacetime with a metric $g_{\mu
\nu }$, and another set $\{\sigma ^{a}\}$ represents the intrinsic
coordinates of the induced geometry, related each other by a transformation
matrix $e_{a}^{\mu }=\frac{\partial x^{\mu }}{\partial \sigma ^{a}}$. Any
point on the spacetime shell trajectory can be endowed with a local basis $%
\{\xi ^{\mu },e_{a}^{\mu }\}$. In this way, the standard definition of
intrinsic metric over the hypersurface $h_{ab}=e_{a}^{\mu }e_{b}^{\upsilon
}g_{\mu \upsilon }$ is recovered in terms of the spacetime metric.

The surface stress tensor $S_{\mu \nu }$ can be obtained from the volume
tensor $T_{\mu \nu }$ as the limit process in the shell thickness,
\begin{equation}
S_{\mu \nu }=\lim\limits_{\varepsilon \rightarrow
0}\int\limits_{-\varepsilon }^{+\varepsilon }T_{\mu \nu }d\xi .
\label{Smunu}
\end{equation}

The projections of Einstein tensor $G_{\mu \nu }$ along the normal
coordinate $\xi $ and the remaining directions over the hypersurface $\Sigma
_{\xi }$ leads to a set of relations

\begin{eqnarray}
\!G_{\xi a} &=&K_{\mid a}-K_{a\mid b}^{b}  \label{Gchia} ,\\
\!2G_{\xi \xi } &=&^{(d-1)}\!R(h)-(K^{2}-K_{ab}K^{ab}) ,\\
\!G_{ab} &=&^{(d-1)}\!G_{ab}+\!\partial _{\xi
}(K_{ab}-h_{ab}K)\!-\!KK_{ab}\!+\frac{1}{2}h_{ab}(K^{2}+K_{cd}K^{cd}).
\label{Gab}
\end{eqnarray}
Here, $^{(d-1)}\!R(h)$ stands for the Ricci scalar of $h_{ab}$ and $K_{ab}$
is the extrinsic curvature of $\Sigma _{\xi }$.

Integrating the eq.(\ref{Gab}) across the shell, the Lanczos equation is
obtained,

\begin{equation}
\gamma _{ab}-h_{ab}\gamma =8\pi \widetilde{G}S_{ab},  \label{lanczos}
\end{equation}
relating the discontinuity of the extrinsic curvature $\gamma
_{ab}=[K_{ab}]=K_{ab}^{+}-K_{ab}^{-}$ and its trace $\gamma $, with the
projected surface stress tensor $S_{ab}$.

From the equation (\ref{Gchia}) we see that the jump across the shell leads
to the continuity equation for $S_{ab}$

\begin{equation}
S_{a\;\mid b}^{\;b}=-\left[ T_{\mu \nu }e_{a}^{\mu }\xi ^{\nu }\right]
=-\left[ T_{a\xi }\right] .  \label{conteq}
\end{equation}

For many cases of physical interest, we consider a perfect fluid
with a bulk stress tensor
\begin{equation}
 T_{\mu \nu }=\left(\sigma u_{\mu }u_{\nu}-\tau (h_{\mu \nu}+u_{\mu }u_{\nu })\right)\delta (X),
\end{equation}
where $u^{\mu }$ is the shell $d-$ dimensional velocity, and
$\sigma $\ and $\tau $ stand for the surface energy density$\ $and
tension, respectively. The delta function represents a matter
distribution localized at the boundary of $\Sigma _{\xi }$.

Even though the Israel treatment for thin shells has proceeded through a
line of increasing success on the understanding of gravitational collapse,
the complexity brought about, for instance, by adding angular momentum, can
turn this method hard to use in practice.

On the other hand, some authors have proposed alternative
approaches, based on the canonical formalism, to rederive the thin
shell dynamics obtained by the Israel method in a number of cases
\cite{h-b,h-k,hajicek,berezin,ansoldi,k-w}. Following this line,
we present a simple method to reproduce the equations of motion
for the radial collapse of thin shells, but that can also be
extended to deal with rotating solutions in three-dimensional
spacetimes.

In the next section, we show that the direct integration of Hamiltonian
constraints provides a complete set of equations equivalent to the ones
obtained from the standard thin shell method.

\section{Hamiltonian Treatment of Thin Shell Collapse}

The Einstein-Hilbert action with cosmological constant in $d$ dimensions is
written as

\begin{equation}
I=-\kappa \int d^{d}x\sqrt{-^{(d)}g}(R-2\Lambda ),  \label{IEH}
\end{equation}
with the constant in front of the gravitational action as $\kappa ={\frac{1}{%
2(d-2)\Omega _{d-2}G}}$ \cite{constants}. The general approach presented
here is equally valid for any value of the constant $\Lambda $. For later
purposes, $\Lambda $ is chosen as $\Lambda =-\frac{(d-1)(d-2)}{2l^{2}}$ in
terms of the AdS radius $l$.

Taking a timelike ADM foliation for the spacetime \cite{ADM}, we write the
line element as

\begin{equation}
ds^{2}=-(N^{_{\bot
}})^{2}dt^{2}+g_{ij}(N^{i}dt+dx^{i})(N^{j}dt+dx^{j}), \label{ds}
\end{equation}
where $g_{ij}$ is the spatial metric and the functions $N^{_{\perp }}$ and $%
N^{i}$ represent the time lapse and the spatial shift, respectively. The
quantities $N^{\perp }$ and $N^{i}$ play the role of Lagrange multipliers of
the constraints $\mathcal{H}_{\bot }\approx 0$ and $\mathcal{H}_{i}\approx 0$%
, so that the gravitational action can be cast in Hamiltonian form

\begin{equation}
I=\int dtd^{d-1}x\left( \pi ^{ij}\dot{g}_{ij}-N^{_{\bot
}}\mathcal{H}_{\bot }-N^{i}\mathcal{H}_{i}\right) ,
\label{action}
\end{equation}
where $\mathcal{H}_{\bot }$ and $\mathcal{H}_{i}$ are given by the formulas

\begin{equation}
\mathcal{H}_{\bot }=-\frac{1}{\kappa \sqrt{g}}\left( \pi _{ij}\pi ^{ij}-%
\frac{1}{\left( d-2\right) }\left( \pi _{i}^{i}\right) ^{2}\right)
+\kappa \sqrt{g}\left( ^{\left( d-1\right) }\!R(g)-2\Lambda
\right) +\sqrt{g}T_{\bot \bot } , \label{hper}
\end{equation}

\begin{equation}
\mathcal{H}_{i}=-2\pi _{i\mid j}^{j}+\sqrt{g}T_{\bot i},  \label{ha}
\end{equation}
in presence of matter fields. Here $^{\left( d-1\right) }\!R(g)$ stands for
the Ricci scalar of the spatial metric $g_{ij}$ and $\pi ^{ij}$ are the
conjugate momenta.

\section{Nonrotating case}

By radial collapse, it is possible adding mass and electric charge to an
already existing \emph{(un)}charged black hole, or producing the black hole
itself over a vacuum state.

In a similar procedure to the ones developed in Refs. \cite{hajicek,ansoldi}
for massive shells, here we show that the integration of the Hamiltonian
constraints along an infinitesimal radial distance on a constant-time slice
reproduces the results of the standard formalism. It will be shown that this
treatment also implies the stability of the event horizon in a generic case.

A spherically symmetric collapsing shell has static interior and exterior
geometries described by Schwarzschild-like coordinates
\begin{equation}
ds_{\pm }^{2}=-N_{\pm }^{2}(r)f_{\pm }^{2}(r)dt_{\pm }^{2}+f_{\pm
}^{-2}(r)dr^{2}+r^{2}d\Omega _{d-2}^{2},  \label{inandout}
\end{equation}
where the matching condition for the time is given by the choice $N_{\pm }=1$%
. The radial coordinate $r$ is continuous across the shell, because it
measures the (intrinsic) area of the shell, that is the same as looked at
from the inside and the outside.

The induced metric of the world tube is simply the one of a $(d-2)-$sphere,

\begin{equation}
ds^{2}=-d\tau ^{2}+R^{2}(\tau )d\Omega _{d-2}^{2}.  \label{intrinsic}
\end{equation}

For spherical symmetry, the Hamiltonian generator $\mathcal{H}_{\bot }$
becomes \cite{HamiltAdS}

\begin{equation}
\mathcal{H}_{\bot }=-\frac{\sqrt{g}}{2\Omega _{d-2}G}\left[ \frac{(d-3)}{%
r^{2}}(1-f^{2})-\frac{(f^{2})^{\prime }}{r}+\frac{(d-1)}{l^{2}}\right] +%
\sqrt{g}T_{\bot \bot } . \label{hperspher}
\end{equation}

We are going to integrate out the constraint $\mathcal{H}_{\bot }=0$ across
a radial infinitesimal length centered in the shell position $r=R(\tau )$ at
a constant time, to express the discontinuities in this component of the
Hamiltonian in terms of $T_{\bot \bot }$. It is straightforward to prove
that all the terms --but the radial derivative-- contribute with a finite
value jump proportional to $\varepsilon $, and they can be indeed ruled out
in the limit $\varepsilon \rightarrow 0$. Thus, the only nonvanishing
contribution comes from the second term

\begin{equation}
-\int\limits_{-\varepsilon }^{+\varepsilon }\frac{(f^{2})^{\prime }}{r}dr=-%
\frac{\triangle (f^{2})}{R}=2\Omega
_{d-2}G\int\limits_{-\varepsilon }^{+\varepsilon }T_{\bot \bot }dr
. \label{discon}
\end{equation}

In the r.h.s. of above equation, $T_{\bot \bot }$ is given by $T_{\bot \bot
}=T^{\mu \nu }n_{\mu }n_{\nu }$, the contraction with the timelike normal
vector in the ADM foliation $n_{\mu }=(-N^{_{\bot }},0,\vec{0})$, that
generates the sequence of constant-time surfaces $\Sigma _{t}$.

On the other hand, adapting another frame to the hypersurface $\Sigma _{\xi
} $, we have a set of coordinates $\{T,X\}$. The tangential axis $T$ that
runs along the velocity $u^{\mu }$ and the direction $X$ goes along the
spacelike normal $\xi ^{\mu }$, in whose origin the delta-function is
located. In this way,

\begin{equation}
T_{\bot \bot }=T^{\mu \nu }n_{\mu }n_{\nu }=\{\sigma u_{\bot }u_{\bot }-\tau
(h_{\bot \bot }+u_{\bot }u_{\bot })\}\delta \left( X\right) .
\label{Tperpergen}
\end{equation}

Without loss of generality, we take a Schwarzchild-like coordinate set $%
x^{\mu }=\{t,r,\phi ^{i}\}$ for outer description of shell collapse. Then,
we can compute $u^{\mu }=\{f^{-2}\alpha ,\dot{R},\vec{0}\}$ and $\xi ^{\mu
}=\{f^{-2}\dot{R},\alpha ,\vec{0}\}$, where the function $\alpha $ is given
by $\alpha =\sqrt{f^{2}+\stackrel{.}{R}^{2}}$. Thus, we obtain an expression
for (\ref{Tperpergen}) as seen from $\{T,X\}$ frame

\begin{equation}
T_{\bot \bot }=\sigma \frac{\alpha ^{2}}{f^{2}}\delta \left( X\right) .
\label{TperperX}
\end{equation}

However, to carry out the integration over $r$, we need to rewrite the
delta-function in the spacetime coordinates system $\{t,r\}$.

\begin{figure}[h]
\begin{center}
\leavevmode \epsfxsize=3in\epsfbox{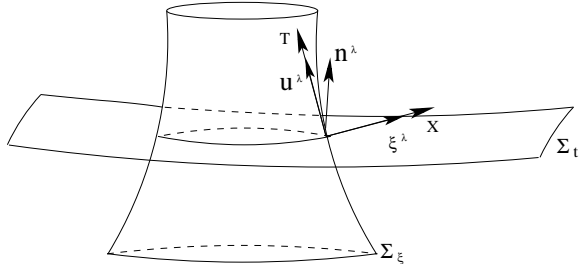}\label{fig1}
\end{center}
\caption{The hypersurfaces $\Sigma _{t}$ and $\Sigma _{\xi }$ are defined by
the normal vectors $n^{\mu }$ and $\xi ^{\mu }$. The intersection between $%
\Sigma _{t}$ and $\Sigma _{\xi }$ is the shell itself at the time $t$.}
\end{figure}
From the Fig. 1, any point on the shell is described by both coordinates
systems as

\begin{eqnarray}
dt &=&u^{t}dT+\xi ^{t}dX , \label{difft} \\
dr &=&u^{r}dT+\xi ^{r}dX.  \label{diffr}
\end{eqnarray}
Integrating along $dr$, on a time-constant ADM slice $\Sigma _{t}$ ($dt=0$),
we get

\begin{equation}
\frac{dr}{dX}=\xi ^{r}-\frac{u^{r}}{u^{t}}\xi
^{t}=\frac{f^{2}}{\alpha } , \label{Jacob}
\end{equation}
and the delta function transforms in such a way that it gets an additional
`relativistic' factor $\delta \left( X\right) =\frac{f^{2}}{\alpha }\delta
\left( r-R\right) $ and the final form for the stress tensor (\ref{TperperX}%
) is

\begin{equation}
T_{\bot \bot }=\alpha \sigma \delta \left( r-R\right) .
\label{TperperR}
\end{equation}

Integrating the above relation, the r.h.s. represents the mean value of the
function $\alpha $ as seen embedded in both inside and outside spacetimes

\begin{equation}
2\Omega _{d-2}G\int\limits_{-\varepsilon }^{+\varepsilon }T_{\bot \bot
}dr=\Omega _{d-2}\sigma \left( \alpha _{+}+\alpha _{-}\right) G.
\label{almost}
\end{equation}

Notice that the tension value $\tau $ does not appear on the right hand side
of (\ref{almost}).

With these simple arguments, this method recovers and extends the
dynamics for radial collapse computed using the thin shell
formalism \cite{cruz-israel,kuchar,ipser-sikivie}

\begin{equation}
-\triangle f^{2}(R)=\left( \Omega _{d-2}R\sigma \right) \left( \sqrt{%
f_{+}^{2}+\stackrel{.}{R}^{2}}+\sqrt{f_{-}^{2}+\stackrel{.}{R}^{2}}\right) G.
\label{MR&MG}
\end{equation}

For three spacetime dimensions, the same formula has been obtained by Steif
and Peleg \cite{steif-peleg} for the gravitational collapse of a dust thin
shell.

Note, as it is well known, that in order to regard (\ref{MR&MG}) as a first
integral of the equation of motion for $R(\tau )$, one needs to specify the
density $\sigma $ as a function of the tension $\tau $. Replacing in the
continuity equation (\ref{conteq}) the expression for $S_{ab}$ and taking
the parallel components to the velocity $u^{a}$ we have

\begin{equation}
(\sigma u^{b})_{\mid b}-\tau u_{\mid b}^{b}=0 , \label{conssigtau}
\end{equation}
that is the relation that provides the conserved quantities in the system.
For example, for coherent dust one has that $R^{d-2}\sigma $ is a constant,
whereas for a domain wall $\sigma $ is a constant. In the first case, the
interpretation of (\ref{MR&MG}) is quite intuitive. For Schwarzchild-AdS
black holes, the function in the metric reads $f^{2}=1-\frac{2GM}{r^{d-3}}+%
\frac{r^{2}}{l^{2}}$, and the term $m=\Omega _{d-2}R^{d-2}\sigma $ is the
rest-frame mass, as seen by an intrinsic observer. Hence, the equation (\ref
{MR&MG}) reduces to
\begin{equation}
\triangle M=\frac{1}{2}\left( \alpha _{+}+\alpha _{-}\right) m,
\end{equation}
that relates the inertial mass to the semisum of the gravitational mass at
each side of the shell. For Minkowskian spacetime, the factor $\alpha $
becomes the special relativity $\gamma $ factor, thanks to the useful
identity $\gamma ^{2}=1+\left( \frac{dR}{d\tau }\right) ^{2}=\left( 1-\left(
\frac{dR}{dt}\right) ^{2}\right) ^{-1}$.

To complete the present picture of radial collapse, it is necessary to
analyze the consistency of the remaining nonvanishing components of the
Hamiltonian.

The angular components of the constraint (\ref{ha}) are identically zero. In
common cases, the condition of spherical symmetry is sufficient to ensure
that the radial constraint $\mathcal{H}_{r}^{(g)}$ vanishes. However, here
it is different from zero because $T_{\bot i}$ is proportional to the radial
velocity. One can expect $\mathcal{H}_{r}$ to be proportional to $\mathcal{H}%
_{\perp }$, since (\ref{hper}) already provides the equation of motion for $%
R(\tau )$. It is interesting to see explicitly that this indeed occurs. The
proof also illustrates again how efficiently one obtains in this approach a
feature already known in the Israel method.

Computing the extrinsic curvature by definition in terms of the Lie
derivative, we get

\begin{equation}
K_{ij}=-\frac{1}{2}\mathcal{L}_{n}g_{ij}=-\frac{1}{2}\partial _{\bot }g_{ij}.
\end{equation}
Here $\partial _{\bot }=n^{\mu }\partial _{\mu }$ defines the derivative
along the ADM timelike normal $n^{\mu }$. This requires the projection of
the vector $n^{\mu }$ on the shell frame, which is decomposed on the basis $%
\{u^{\mu },\xi ^{\mu }\}$ as $n^{\mu }=au^{\mu }+b\xi ^{\mu }$, on the
intersection between the shell hypersurface $\Sigma _{\xi }$ and the
constant-time slice $\Sigma _{t}$. Projecting between the frames, we obtain
the coefficients $a=f^{-1}\alpha $ and $b=-f^{-1}\dot{r}$, which allows to
express the normal derivative as

\begin{equation}
\partial _{\bot }=\frac{\alpha }{f}\frac{\partial }{\partial \tau }-\frac{%
\dot{r}}{f}\frac{\partial }{\partial X}.  \label{normalder}
\end{equation}
Here, we have used the definitions $u^{\mu }\partial _{\mu }=\frac{\partial
}{\partial \tau }$ and $\xi ^{\mu }\partial _{\mu }=\frac{\partial }{%
\partial X}$.

The metric $g_{ij}$ has no dependence on $X$, because the coordinate $X$ can
be always set to zero over $\Sigma _{\xi }$. Then, the explicit form for the
extrinsic curvature $K_{ij}$ as a proper time derivative of $g_{ij}$ is
\begin{equation}
K_{ij}=-\frac{\alpha }{2f}\frac{\partial g_{ij}}{\partial \tau }.
\end{equation}

Imposing the constraint over the radial component of eq. (\ref{ha}) leads to

\begin{equation}
-2\pi _{r\;\mid j}^{\ j}+\sqrt{g}T_{\bot r}=0 ,
\end{equation}
where $\pi ^{ij}$are obtained by means of the above formula for
$K_{ij}$ calculated with the spatial metric $g_{ij}$ of ADM
foliation.

Computing the stress tensor in terms of velocity and intrinsic metric, and
using the Jacobian of the basis change, produces

\begin{equation}
T_{\bot r}=\frac{f^{2}}{\alpha }S_{\bot r}\delta (r-R)=-\frac{\dot{r}}{f}%
\sigma \delta (r-R) , \label{Tperr}
\end{equation}
resulting in the equation

\begin{equation}
\frac{d}{dr}\left( \alpha \dot{r}\right) =-\Omega _{d-2}r\dot{r}\sigma
\delta (r-R)G.
\end{equation}
The integration of this relation gives the discontinuity in the function $%
\alpha $ across the shell

\begin{equation}
\alpha _{+}-\alpha _{-}=-\Omega _{d-2}R\sigma G.  \label{disalpha}
\end{equation}

In the context of standard thin shell formalism, this equation comes from
the discontinuity in the normal acceleration across the hypersurface $\Sigma
_{\xi }$. However, this does not stand for an independent relation from the
energy conservation law (\ref{MR&MG}), since it can also be recovered
multiplying that equation by $(\alpha _{+}-\alpha _{-})$.

The equation (\ref{MR&MG}) has clearly a limited range of validity
in the Schwarzschild-like radial coordinate $R$, since $-\left(
\Delta f^{2}\right) $ must be strictly positive. For instance, for
radial collapse of a massive thin shell, the l.h.s. of this
relation is just the difference of the outer solution mass respect
AdS spacetime, that is positive for all the solutions of physical
interest. In a more general case, there might be a radial position
where $\Delta f^{2}$ vanishes. However, the same formula, written
in the form of eq. (\ref{disalpha}) tells us that the shell must
bounce back before this happens, because for $\dot{R}=0$

\begin{equation}
f_{+}-f_{-}=-\Omega _{d-2}R\sigma G.
\end{equation}
As it is well known, the analysis of the shell motion can be carried out
until the point where $f_{+}^{2}=0$. The change in the signature of the
metric in the outer side leads to an inconsistency in the matching
conditions on the shell.

Whereas the previous discussion in general imposes a lower bound for $R$,
the positive definiteness of the functions $\alpha _{\pm }$ makes the
analysis to break down beyond a critical radius, for instance, in the black
hole formation from a domain wall collapse, discussed below.

\section{Radial Collapse in Three Dimensions Spacetime}

For simplicity, we will focus ourselves on the problem of black hole
creation in (2+1) dimensions, setting the inner solution as AdS spacetime ($%
M=-1$).

\subsection{Coherent Dust Shell Collapse}

For presureless dust, $m=2\pi R\sigma $ is a constant of motion.
In this case, already studied in the Ref. \cite{steif-peleg}, we
have that (\ref{MR&MG}) takes the form

\begin{equation}
M+1=\frac{1}{2}(\alpha _{+}+\alpha _{-})m , \label{dustcollapse}
\end{equation}
with $M+1\geq 0$. For a given value of $m$, the collapse comes
from the radial speed expression

\begin{equation}
\dot{R}^{2}=\left( \frac{a^{2}}{16m^{2}}-1\right) -R^{2} ,
\label{Rdotdust}
\end{equation}
with $a=m^{2}+4(M+1)$, gravity constant $G=\frac{1}{2}$ and AdS radius $l=1$.

Its analysis leads to a confined motion for the dust ring, because the dust
ring cannot be located beyond the turning point $R_{0}^{2}=\left(
a^{2}-16m^{2}\right) /16m^{2}$. This distance turns out to be greater than
the black hole horizon for any outer solution with $M>0$.

Depending on initial velocity and position, either a naked singularity ($%
-1<M<0$) or a black hole ($M>0$) can be formed from this radial
gravitational collapse process, as stated by Peleg and Steif. For negative
mass solutions, there exists a critical shell mass $m=2\sqrt{\left| M\right|
}+1$ below which the motion is impossible in the whole space. Apart from
this condition, the analysis of the effective potential does not constitute
a physical impediment to prevent the creation of a naked singularity in the
black hole mass gap $(-1,0)$. However, as we shall see in Section 6, the
introduction of a however small amount of rotation gets rid of the naked
singularities.

\subsection{Closed Fundamental String Collapse}

The radial collapse of a fundamental string can also generate a black hole
(or naked singularity) as the external configuration starting up from AdS
spacetime as the interior solution, for certain initial conditions.

In this case, it is more useful to analyze the equation of the radial
acceleration, rather than its first integral (\ref{MR&MG}). One obtains, by
differentiation of (\ref{MR&MG})

\begin{equation}
\ddot{R}=-R-\pi \frac{\alpha _{+}\alpha _{-}}{R} ,
\label{acdomain}
\end{equation}
which implies that $\ddot{R}<0$ and therefore there is no bounce, because
the functions $\alpha _{\pm }$ are always positive. Hence, the gravitational
collapse is unavoidable for any shell density $\sigma $ and black hole mass $%
M$.

Another interesting feature is that, just as it happens in $3+1$ as pointed
in \cite{ipser-sikivie}, due to the particular form of the functions $f_{\pm
}^{2}$ present in the metric, the constraint (\ref{disalpha}) is violated if
$R>R_{\max }=\left( \pi ^{2}\sigma ^{2}-1\right) ^{-1/2}$, with $R_{\max }$
as the maximum value of $R$ (the value for which $\dot{R}=0$). The existence
of this bound for the radial coordinate makes the dynamical analysis unable
to treat the cases where this critical radius is located within the event
horizon $r_{+}$ and, therefore, the system is already collapsed.

From this consideration, it comes that there exist only an allowed interval
in the mass spectrum for the exterior solution with a given density $\sigma $

\begin{equation}
M=2\pi \sigma R_{\max }\sqrt{1+R_{\max }^{2}}-\left( \pi \sigma \right)
^{2}R_{\max }^{2}-1.  \label{massdomain}
\end{equation}
In the same way as in the (3+1)-dimensional counterpart, this process cannot
create black hole solutions beyond that mass range, where too large
spherical walls are already collapsed inside their corresponding
Schwarzschild radius \cite{ipser-sikivie}.

\subsection{General Case}

The existence of the equation of state determines the nature of the
collapsing matter, ranging from coherent dust ($\tau =0$) to a domain wall ($%
\tau =\sigma $). Interpolating between these cases, we can set a parameter $%
\alpha $, such that $\tau =\alpha \sigma $.

Choosing the commoving frame in the equation (\ref{conssigtau}) and
introducing the value of the tension, we can see that the density satisfies
at any time the relation

\begin{equation}
\sigma =C_{0}R^{\alpha -1} , \label{densitygen}
\end{equation}
where $C_{0}$ is a constant throughout the motion.

For an even more general dependence of the tension $\tau $, we can always
write the equation of motion as

\begin{equation}
\ddot{R}=-R-\pi \frac{\alpha _{+}\alpha _{-}}{R}\frac{\tau }{\sigma }.
\label{generalacc}
\end{equation}

Provided $\tau \geq 0$, eq. (\ref{generalacc}) tells us that
$\ddot{R}$ is always negative. As a consequence, the shell
accelerates inwards and it will always collapse to either a black
hole or a naked singularity, depending on the initial conditions.

\subsection{Electrically Charged Solutions}

Electrically charged solutions are obtained supplementing the
Einstein-Hilbert action (\ref{IEH}) by the Maxwell term

\begin{equation}
I_{Maxwell}=\frac{1}{4\epsilon \Omega _{d-2}}\int d^{d}x\sqrt{-^{(d)}g}%
F_{\mu \nu }F^{\mu \nu },  \label{Maxwell}
\end{equation}
in an arbitrary dimension $d$. The constant $\epsilon $ can be written in
terms of the vacuum permeability as $\epsilon =\epsilon _{0}/\Omega _{d-2}$.

For an static, spherically symmetric $\emph{Ansatz}$, the
Reissner-Nordstr\"{o}m-AdS black hole metric appropiately describe the
geometry of both inner and outer regions of spacetime

\begin{equation}
f_{\pm }^{2}=1+\frac{r^{2}}{l^{2}}-\left( \frac{2GM_{\pm }}{r^{d-3}}-\frac{%
\epsilon G}{d-3}\frac{Q_{\pm }^{2}}{r^{2(d-3)}}\right) ,
\label{RNmetric}
\end{equation}
where the shell carries an electric charge $q=Q_{+}-Q_{-}$.

The general form of eq. (\ref{MR&MG}) that governs the radial
collapse in any dimension, remains the same in this case because
the electromagnetic stress tensor does not contribute to the
Hamiltonian component $\mathcal{H}_{\bot } $. Therefore, the
equation of motion becomes

\begin{equation}
\Delta M-\frac{\epsilon }{2\left( d-3\right) }\frac{\Delta Q^{2}}{R^{d-3}}=%
\frac{1}{2}\left( \Omega _{d-2}R^{d-2}\sigma \right) \left( \alpha
_{+}+\alpha _{-}\right) , \label{motionRND}
\end{equation}
that recovers the thin shell dynamics studied in
\cite{cruz-israel,kuchar} for the 4-dimensional case.

In $\left( 2+1\right) $ dimensions, the solution corresponding to an
electrically charged static black hole was first presented in the reference
\cite{BTZ} as the three-dimensional counterpart of the R-N black hole. The
metric contains a logarithmic dependence on the radial coordinate,

\begin{equation}
f^{2}=r^{2}-M-\frac{1}{4}Q^{2}\ln r^{2} , \label{3Q}
\end{equation}
with the constant $\epsilon $ and the cosmological length $l$ set equal to
unity.

From the analysis of this function, the condition for the existence of
extremal black holes is

\begin{equation}
M=\frac{Q^{2}}{4}\left[ 1-\ln \frac{Q^{2}}{4}\right] ,
\label{extremalQ}
\end{equation}
that is the curve that separates black holes configurations from naked
singularities in the plane $(M,Q)$. If the electric charge is large enough,
there exist black hole solutions for arbitrarily negative values for the
mass.

In order to study the creation of charged black holes over a vacuum state,
we set the inner solution as $AdS$ spacetime, with $f_{-}^{2}=\left(
1+r^{2}\right) $. With a dust shell carrying a total mass $m=2\pi R\sigma $,
the equation (\ref{MR&MG}) becomes

\begin{equation}
M+1+\frac{Q^{2}}{4}\ln R^{2}=\frac{m}{2}(\alpha _{+}+\alpha _{-})
, \label{motion3Q}
\end{equation}
and the exterior mass and charge as $M_{+}=M$ and $Q_{+}=Q$, respectively.
The l.h.s of the above expression must be positive in order to ensure the
validity of the treatment in this coordinates set, and therefore, it imposes
a lower bound for the radial coordinate $R^{2}>e^{-4(M+1)/Q^{2}}$. It can be
proved that this quantity is larger than the inner horizon $r_{-}$ for any
charged black hole and its existence is only relevant in the context of
naked singularities creation, discussed below.

The radial velocity for this case is obtained by quadrature and takes the
form

\begin{equation}
\dot{R}^{2}=-(R^{2}+1)+\frac{1}{16m^{2}}\left( a+b\ln R^{2}\right)
^{2}, \label{Rdot3Q}
\end{equation}
with the constants $a$ and $b$ defined in terms of the parameters of the
solution as $a=m^{2}+4(M+1)$ and $b=Q^{2}$. A quick analysis of the function
shows that there must necessarily be a turning point as we move towards
infinity (in the most general case there could be even two more). To find
the local maximal and minimal points $\bar{R}$ for the effective potential
one solves the trascendental equation

\begin{equation}
\frac{8m^{2}}{b}\bar{R}^{2}=a+b\ln \bar{R}^{2}.
\end{equation}
Keeping $a$ and $m$ to a fixed value, the limit of $b\rightarrow 0$ produces
that both intersection points move to $\bar{R}_{1,2}^{2}\rightarrow 0$ . On
the contrary, if the limit in the parameter $b\rightarrow \infty $ is taken,
the extremal points are shifted to $\bar{R}_{1}^{2}\rightarrow 1$ and $\bar{R%
}_{2}^{2}\rightarrow \infty $.

An inflection point exists, at the position $\bar{R}^{2}=\frac{b^{2}}{8m^{2}}
$ when the parameters satisfy $a=b\left( 1-\ln \frac{b^{2}}{8m^{2}}\right) $%
. The corresponding radial velocity at that point is always purely
imaginary. This relation represents a critical value for $a$ and $b$, that
permits the existence of local extremal points in the curve for $a$ over
that value. For values of $a$ below the one given by the equality, there is
neither local maximum nor minimum, the curve is monotonously decreasing and
the only turning point is immersed in the zone where the equation of motion
is no longer valid.

Another critical situation is represented by a static thin shell, where the
dust ring has been put in a fixed radial position
\begin{equation}
R_{*}^{2}=\frac{b}{8m^{2}}\left[ b+\sqrt{16m^{2}+b^{2}}\right],
\end{equation}
and it is not able to collapse or expand. For this particular case, the
solution parameters satisfy the relation

\begin{equation}
a=b\left( 1-\ln \left( \frac{b}{8m^{2}}\left[ b+\sqrt{16m^{2}+b^{2}}\right]
\right) \right) +\sqrt{16m^{2}+b^{2}}.
\end{equation}
Note that if we take the total mass of the shell as $m^{2}=b+4$, the process
creates an extremal black hole with a ring standing still at the event
horizon $R_{+}^{2}=b/4$.

However, it is important to stress that this situation represents
just a critical case in the extremal black holes formation, as
there are many different sets of initial conditions that also
generate them. From this perspective, extremal black holes cannot
be regarded as `fundamental' objects, because eq. (\ref{Rdot3Q})
allows their creation from the dynamic process depicted in this
section.

Finally, from the analysis of the effective potential (\ref{Rdot3Q}) we
conclude that a charged spherically symmetric shell cannot collapse to form
a naked singularity in three dimensions. It is worthwhile to stress that, in
spite of the different form of the charged black hole metric and the
extremality condition derived from it, this property is also found in the
four-dimensional case \cite{boulware}.

\section{Rotating Black Hole Solutions in Three Dimensions}

A different case is represented by the rotating black hole in a $(2+1)-$%
dimensional spacetime. This time, the line element possesses a shift along
the angular direction, responsible for the existence of two horizons and an
ergosphere \cite{BTZ}, in an analogous way to the Kerr metric in $(3+1)$
dimensions

\begin{equation}
ds^{2}=-N^{2}f^{2}dt^{2}+f^{-2}dr^{2}+r^{2}(N^{\phi }dt+d\phi
)^{2}, \label{BTZrot}
\end{equation}
where

\begin{eqnarray}
f^{2} &=&-M+\frac{r^{2}}{l^{2}}+\frac{J^{2}}{4r^{2}}, \\
N^{\phi } &=&-\frac{J}{2r^{2}}+N^{\phi }\left( \infty \right), \\
N &=&N\left( \infty \right).
\end{eqnarray}
The residual arbitrariness constitutes the choice of\textbf{\ }$N$ at
infinity, which is usually set as $N\left( \infty \right) =1$ and the
angular shift $N^{\phi }$. For this case, we will choose $N^{\phi }(R(\tau
))=0$, that represents a null angular velocity on the shell at every time,
and simply corresponds to a reparametrization in the angular variable. In
this form it is possible to attain suitable matching conditions on the
shell, for instance, for a static internal solution.

The rotating solution possesses the same isometries as the static one, the
Killing vectors $\partial _{t}$ and $\partial _{\phi }$. This makes sensible
the vector basis choice for both outside and inside spaces in a similar way
as in the previous case. Therefore, the projection of the 3-velocity along
the basis $\left\{ n,\frac{\partial }{\partial r},\frac{\partial }{\partial
\phi }\right\} $ can be cast in the form

\begin{equation}
u^{\mu }=\frac{\alpha }{f}n^{\mu }+\dot{r}\left( \frac{\partial }{\partial r}%
\right) ^{\mu }+u^{\phi }\left( \frac{\partial }{\partial \phi
}\right) ^{\mu } , \label{urot}
\end{equation}
and the normal vector $\xi ^{\mu }$ in terms of the same orthogonal set

\begin{equation}
\xi ^{\mu }=\frac{\dot{r}}{f\gamma }n^{\mu }+\frac{\alpha }{\gamma
}\left( \frac{\partial }{\partial r}\right) ^{\mu } ,
\label{xirot}
\end{equation}
with the angular velocity defined as $u^{\phi }=\frac{d\phi }{d\tau }.$

The functions $\alpha $ and $\gamma $ have the explicit expressions
\begin{equation}
\alpha ^{2}=f^{2}+\dot{r}^{2}+f^{2}r^{2}(u^{\phi })^{2},
\end{equation}
and
\begin{equation}
\gamma ^{2}=1+r^{2}(u^{\phi })^{2}.
\end{equation}

It is useful to define a new time coordinate
\begin{equation}
d\lambda =\sqrt{1+r^{2}\left( \frac{d\phi }{d\tau }\right) ^{2}}d\tau ,
\end{equation}
that corresponds to the proper time measured by an observer in radial
collapse. In this way, the angular velocity can be expressed as

\begin{equation}
\Omega =\frac{d\phi }{d\lambda }=\frac{\dot{\phi}}{\sqrt{1+r^{2}\dot{\phi}%
^{2}}} , \label{Omega}
\end{equation}
that, in turn, permits to write down the time variable and the angular
velocity as

\begin{eqnarray}
d\lambda &=&\gamma d\tau ,\\
\dot{\phi} &=&\gamma \Omega,
\end{eqnarray}
in an analogous way to special relativity, using the (dilation) relativistic
factor
\begin{equation}
\gamma =\frac{1}{\sqrt{1-r^{2}\Omega ^{2}}}.
\end{equation}

Once more, the equation $T_{\mu \nu }=\left(\sigma u_{\mu }u_{\nu
}-\tau (h_{\mu \nu }+u_{\mu }u_{\nu })\right)\delta \left(
X\right)$ provides the shell stress tensor, with a delta
distribution located at the origin of $X$ axis, along $\xi ^{\mu
}$ direction. Computing the relevant components and expressing
them in terms of the normal time $\lambda $

\begin{eqnarray}
T_{\bot \bot } &=&\frac{\alpha ^{2}}{f^{2}}\left\{ \gamma
^{2}\sigma -\tau (\gamma ^{2}-1)\right\} \delta \left( X\right) ,
\label{Tperperrot}
\end{eqnarray}
\begin{eqnarray}
T_{\bot \phi } &=&-\frac{\alpha }{f}\gamma ^{2}r^{2}\Omega (\sigma -\tau
)\delta \left( X\right) .  \label{Tper2rot}
\end{eqnarray}
In this case, the function $\alpha $ has been defined as

\begin{equation}
\alpha ^{2}=f^{2}+\left( \frac{dr}{d\lambda }\right) ^{2}.
\end{equation}

Performing the required change of variable to integrate out in the radial
coordinate $r$, the Jacobian $\frac{dr}{dX}=\frac{f^{2}}{\alpha }$ remains
unchanged with the new time definition. It is clear that the whole procedure
matches the radial collapse case when $\Omega =0$.

Again the discontinuity in $\mathcal{H}_{\bot }$\textbf{\ }are caused only
by one term in $^{\left( 2\right) }R$\textbf{, }because all other terms
represent finite jumps in a null-measure interval.\textbf{\ }Thus,\textbf{\ }%
the equation \textbf{(}\ref{MR&MG}\textbf{)} undergoes a change, due to the
different form of $T_{\bot \bot }$, and becomes

\begin{equation}
-\triangle (f^{2})=\pi R(\alpha _{+}+\alpha _{-})\left\{ \gamma ^{2}\sigma
-\tau (\gamma ^{2}-1)\right\} .  \label{deltaf2}
\end{equation}
The fact that $\gamma $ has the same value at each side of the shell is a
direct consequence of the junction conditions imposed on the shell position $%
r=R(\tau )$.

The direct integration of the angular component of Hamiltonian\textbf{\ }$%
\mathcal{H}_{\phi }$ is possible considering the only nonvanishing component
of the gravitational momentum $\pi _{\phi }^{\;r}=p(r)/2\pi $. This\textbf{\
}contributes a the difference in the angular momentum\textbf{\ }$\triangle J$%
, coming from (\ref{Tper2rot}), given by

\begin{equation}
-2\Delta p=\triangle J=2\pi \gamma ^{2}R^{3}\Omega (\sigma -\tau ).
\label{deltaJ}
\end{equation}

The equivalent of the eq. (\ref{disalpha}) can be obtained from (\ref{deltaf2}%
) repeating the same analysis depicted in Sec.4,

\begin{equation}
\alpha _{+}-\alpha _{-}=-\pi R\sigma \left\{ \gamma ^{2}\sigma -\tau (\gamma
^{2}-1)\right\} .  \label{disalphaang}
\end{equation}
This is an useful version of the equation of motion for the study of the
dynamical interval in the radial coordinate.

These equations provide the starting point for the analysis of the collapse
of a rotating shell. In the cases shown below, the extremal values for shell
energy density and tension are explicitly developed. We will focus in the
process of black hole formation onto a `vacuum' inner solution (AdS
spacetime).

\subsection{Domain Wall ($\sigma =\tau $)}

A rotating shell with a tension equal to the mass density represents a
singular case of the equations of motion governing the collapse dynamics.
From the equation (\ref{deltaJ}) we see that the contribution to the angular
momentum is vanishing for a collapsing domain wall. This was geometrically
expected due to the fact that for $\sigma =\tau $, this object can be
obtained from the Nambu-Goto action for a fundamental string. The Poincar%
\'{e} symmetry defines an angular momentum tensor that is identically
vanishing for a perfectly circular rotating string.

This result states the impossibility to generate rotating
solutions with this `fundamental object'. Furthermore, the
condition imposed on the eq. (\ref{deltaf2}) reproduces the same
expression (\ref{MR&MG}) as for the nonrotating domain wall
collapse, for an observer falling radially with the shell.

\subsection{Dust Shell}

The collapse of a presureless shell represents a system of particles
travelling inwards with no mutual interaction. Thus, the path of every
infinitesimal piece of matter is given by the geodesics in an external
gravitational field, spinning around the radial potential because of the
initial angular velocity.

For this case, the equations (\ref{deltaf2}) and (\ref{deltaJ}) that set the
change in the parameters between AdS and the outer spacetime, take the form

\begin{equation}
M+1-\frac{J^{2}}{4R^{2}}=\pi \sigma R(\alpha _{+}+\alpha
_{-})\gamma ^{2}, \label{wa1}
\end{equation}
for the energy conservation, and

\begin{equation}
J=2\pi \gamma ^{2}R^{3}\sigma \Omega , \label{wa2}
\end{equation}
for the angular momentum. The description here is from the frame of an
observer falling radially with the shell (non rotating), that measures a
time $\lambda $. The equation (\ref{conssigtau}) gives the conservation of
the total mass, enlarged in a $\gamma $ factor respect to the commoving
(rest) frame

\begin{equation}
2\pi \gamma R\sigma =m.  \label{massang}
\end{equation}

Replacing the latter expression in (\ref{wa2}) allows us to obtain the
angular velocity

\begin{equation}
\Omega =\frac{\pm J}{R\sqrt{J^{2}+m^{2}R^{2}}} , \label{omegaJ}
\end{equation}
where the plus (minus) sign stands for the shell rotating
(counter)clockwise; and the explicit form for the relativistic $\gamma $
factor

\begin{equation}
\gamma =\frac{1}{\sqrt{1+\frac{J^{2}}{m^{2}R^{2}}}}.  \label{gammaJ}
\end{equation}
Finally, inserting all these results in (\ref{wa1}), the radial velocity $%
\stackrel{\circ }{R}=\frac{dR}{d\lambda }$ as a function of the solution
parameters and the radial coordinate is

\begin{equation}
\stackrel{\circ
}{R}^{2}=\frac{a^{2}R^{2}}{16(m^{2}R^{2}+J^{2})}-(R^{2}+1),
\label{poteff}
\end{equation}
with the constant $a$ again defined as $a=m^{2}+4(M+1)$.

The maximum value of the above function is found to be $R_{\max
}^{2}=J(a-4J)/4m^{2}$. A quick analysis of the effective potential shows
that the shell cannot reach the origin $R=0$, nor the infinity, confining
the motion between two turning points. In order to ensure that these turning
points do not coalesce --when they indeed exist-- the maximum value for $%
\stackrel{\circ }{R}$

\begin{equation}
\stackrel{\circ }{R}_{\max }^{2}=\frac{(a-4J)^{2}}{16m^{2}}-1,
\end{equation}
must be greater than zero. Therefore, for the motion to exist at all, the
parameters must satisfy the condition $\frac{m}{2}>\sqrt{J-M}+1$ or $0<\frac{%
m}{2}<\sqrt{J-M}-1$ for the possible creation of a naked singularity ($J>M$%
). However, as the shell does not disappeared beyond an event
horizon, necessarily the bounce is produced for any value of the
initial conditions. Thus, the dust ring cannot generate the naked
singularity at the origin. The presence of angular momentum
provides a `centrifugal barrier' that is not infinite as in the
Keplerian case, and whose effect is clear when we put eq.
(\ref{poteff}) into the form
\begin{equation}
\stackrel{\circ }{R}^{2}=V_{eff}(J=0)-\frac{a^{2}J^{2}}{%
16m^{2}(m^{2}R^{2}+J^{2})},
\end{equation}
where $V_{eff}(J=0)$ corresponds to the r.h.s of eq. (\ref{Rdotdust}).

In view of the above result, we can reinterpretate the only case in three
dimensions where it was possible to form naked singularities: the radial
collapse of a massive shell onto AdS vacuum. Because of the existence of a
mass gap between AdS and the M=0 black hole, the outer solution can have a
negative mass even for a shell with $\sigma >0$. However, this case is
somehow ill-defined because the particles would need to free-fall with
infinite precision along the radial direction. Any angular perturbation in
the initial condition would prevent the shell to reach the origin.

In turn, outer black hole solutions ($J<M$) are created for any value of the
shell mass $m$, since the smallest turning point is always inside the
horizon $r_{+}$. The time-evolution is completely determined once the
initial conditions are set. In particular, for a collapsing shell starting
from zero radial velocity at a distance $R=R_{0}$, we obtain the expression
for the mass of the external solution

\begin{equation}
M=\frac{2\sqrt{(R_{0}^{2}+1)(m^{2}R_{0}^{2}+J^{2})}}{R_{0}}-\left( \frac{%
m^{2}}{4}+1\right) .  \label{Mrot}
\end{equation}

For extremal black holes, there is no restriction in the total mass of the
collapsing ring, either. The limit case is represented by the situation
where both turning points coalesce. Because the shell mass must be $m=2$,
the ring is orbiting steadily at a fixed radius $R^{2}=\frac{J}{2}$, the
radius corresponding to an extremal black hole horizon. As a consequence,
the shell dynamics sees no objection to the formation of extremal black
holes from a collapse process with an appropiate set of initial conditions,
in a similar way as in the charged black hole creation.

\section{Conclusions}

Apart from the relative ease this alternative treatment reproduces and
extends the dynamics for collapsing thin shells obtained by the Israel
method, this formalism presents a few additional interesting features,
especially because of the general statements that can be derived from.

The geometrical scheme applied in the derivation of the formula (\ref{MR&MG}%
) --and the corresponding version in the rotating case-- permits to write
them generically in terms of the change in the geometry through the shell $%
\triangle f^{2}$ and not explicitly in terms of any particular solution
parameters ($M,J,Q,\Lambda $, etc.). Even subtle, it is precisely this
difference which generalizes the method, opening the possibility of dealing
with a number of interesting cases: from black hole creation --as presented
in this letter-- to thin shells collapse over an existing black hole, as
also possible to extend for higher dimensional spacetimes.

In $(2+1)$ dimensions, a direct consequence of equations (\ref{MR&MG}) and (%
\ref{deltaf2}) is the well-known thermodynamical law stating that the
horizon area always grows. This can be derived from the energy conservation
law for both nonrotating and rotating cases as follows: let assumed that a
thin shell of physical matter --satisfying $\sigma >\tau >0$-- is dropped
over an already existing black hole configuration. Therefore, the l.h.s. in (%
\ref{MR&MG}) and (\ref{deltaf2}) is strictly positive for any value of $R$,
that is, $f_{+}^{2}(R)<f_{-}^{2}(R)$. For the interior black hole, there
exists an even horizon $R_{+}^{(in)}$ such that $f_{-}^{2}\left(
R_{+}^{(in)}\right) =0$, and the function $f_{+}^{2}$ must be negative for
the same position. Hence, this last function should vanish at a larger
distance than the inner horizon $R_{+}^{(in)}$. The condition impossed mean
that any mechanical perturbation would not move faster than the speed of
light around the shell and it is equivalent to the usual \emph{dominant}
energy condition in cosmology (see, $\emph{e.g.}$, Ref.\cite{waldGR}). The
former argument is also valid for radial collapse in higher dimensions.

Another general consequence, regarding naked singularity formation from the
collapse of a thin shell over a black hole interior solution, can be made
from the analysis of equation (\ref{MR&MG}).

Let the set of parameters be such that the collapse will turn the inner
black hole solution into a naked singularity, as seen by a distant external
observer. For example, we can imagine a near-extremal electrically charged
black hole and a dust shell carrying more charge $q$ than proper mass $m$.
For this system, we have an inner event horizon $R_{+}^{(in)}$, such that $%
f_{-}^{2}\left( R_{+}^{(in)}\right) =0$, whereas the exterior function $%
f_{+}^{2}(R)$ is positive throughout the space. Roughly speaking, if the
shell does not gather enough speed during the collapse, it will not become
massive enough to prevent the formation of a naked singularity. Furthermore,
from the shell dynamics we know that a matter sphere released with certain
speed is equivalent to one dropped from the rest at another distance. Then,
in principle, it might be always possible to find a set of initial
conditions to destroy the black hole configuration.

However, eq. (\ref{MR&MG}) expresses that by the time the shell
has reached the black hole horizon $R_{+}^{(in)}$, the
conservation of energy has already been violated. In addition, the
shell must have bounced before, because for $\dot{R}=0$ at the
horizon

\begin{equation}
f_{+}\left( R_{+}^{(in)}\right) =-\Omega _{d-2}R\sigma G,
\end{equation}
in open contradiction with the fact we have an external naked singularity. A
similar argument can be developed for the rotating case in 3 dimensions,
stating the impossibility of turning black holes into naked singularities by
throwing thin shells of physical matter over.

The previous reasoning cannot be repeated \emph{verbatim} in the case of
naked singularity formation over an empty space. Nevertheless, as we
discussed in the corresponding sections, the absence of a horizon and the
explicit form of the metric for the cases with angular momentum and electric
charge, prevents the shell to reach the origin.

Finally, the Hamiltonian formalism for the collapse of thin shells developed
in this paper can be applied to create magnetic black holes in three
spacetime dimensions \cite{olea}. It can also be extended to deal the
problem of gravitational collapse in gravity theories with higher powers in
the curvature \cite{CdCS}

\section{Acknowledgments}

We wish to thank Prof. C. Teitelboim for having proposed the problem to us
and for numerous enlightening discussions. Institutional support to Centro
de Estudios Cient\'{\i}ficos (CECS) from Empresas CMPC is acknowledged. CECS
is a Millennium Science Institute and is funded in part by grants from
Fundaci\'{o}n Andes and the Tinker Foundation. This work was partially
funded by Ministerio de Educaci\'{o}n through MECESUP grant FSM 9901 and the
grant 3030029 from FONDECYT.


\begin{thebibliography}{99}
\bibitem{constants}  The choice of the factor in front of the action differs
from the usual $-(16\pi \widetilde{G})^{-1}$ for all spacetime dimensions,
where $\widetilde{G}$ is the Newton constant in 4 dimensions. The use of
this convention simplifies the form of the black hole metric in higher
dimensions. The gravity constants are related by the formula $G=\left[ 8\pi
/(d-2)\Omega _{d-2}\right] \tilde{G}$, where $\Omega _{p}=2\pi
^{(p+1)/2}/\Gamma ((p+1)/2)$ stands for the volume of $S^{p}$.
\end{thebibliography}
\end{document}